\documentclass[epj]{svjour}
\usepackage{graphics}
\usepackage{amsmath}
\usepackage{color}
\begin{document}
\title{First-principles particle simulation and Boltzmann equation analysis of Negative Differential Conductivity and Transient Negative Mobility effects in xenon}


\author{Zolt\'an Donk\'o \inst{1}, Nikolay Dyatko \inst{2}
}                     
%
%
\institute{Institute for Solid State Physics and Optics, Wigner Research Centre for Physics, Hungarian Academy of Sciences, Konkoly-Thege Mikl\'os str. 29-33, H-1121 Budapest, Hungary \and State Research Center of Russian Federation Troitsk Institute for Innovation and Fusion Research, Troitsk, Moscow
142190, Russia}
\date{Received: date / Revised version: date}
%
\abstract{
The Negative Differential Conductivity and Transient Negative Mobility effects in xenon gas are analyzed by a first-principles particle simulation technique and via an approximate solution of the Boltzmann transport equation (BE). The particle simulation method is devoid of the approximations that are traditionally adopted in the BE solutions in which (i) the distribution function is searched for in a two-term form, (ii) the Coulomb part of the collision integral for the anisotropic part of the distribution function is neglected, (iii) Coulomb collisions are treated as binary events, and (iv) the range of the electron-electron interaction is limited to a cutoff distance. The results obtained from the two methods are, for both effects, in good qualitative agreement, small differences are attributed to the approximations listed above.
\PACS{
      {52.25.Dg}{Plasma kinetic equations}   \and
      {52.65.-y}{Plasma simulation}
     } 
} 
\maketitle
\section{Introduction}
\label{intro} 

Charged particle swarms have been investigated extensively during the past decades both experimentally and theoretically, see e.g. \cite{ZP,Florian,Loff,Pinhao,Trunec,White,p3,white2009}. The experiments, in which the transport of a (usually low density) particle cloud is studied, yield transport coefficients, like the drift velocity, diffusion coefficients, as well as reaction (excitation and ionization) rate coefficients. In theoretical / computational studies electron transport phenomena can be addressed using kinetic approaches \cite{Pinhao}, as well as fluid analysis \cite{p3}. The kinetic approaches yield, as a principal result, the central quantity of kinetic theory, the {\it velocity distribution function} VDF, $f({\bf r},{\bf v},t)$, from which the transport coefficients can be derived. Such calculations can also assist improving the accuracy of the cross section sets, e.g. \cite{cs-adjust}.

The computational methods of kinetic theory can be split into two major groups. {\it Particle-based methods} trace a large number of individual particles in the external field(s) \cite{MC} and via proper sampling schemes they allow the construction of $f({\bf r},{\bf v},t)$ and the computation of swarm transport parameters. {\it Boltzmann equation} (BE) approaches \cite{BE} use different types of expansions of the VDF and solve the set of the resulting differential equations. Once the VDF is found, the transport parameters can readily be derived \cite{BE}. The most widespread approach, the ``two-term approximation'', retains only the first two terms in the Legendre polynomial expansion of the VDF, and is therefore limited to scenarios where the VDF has a small anisotropy. Solvers based on this approximation are available even as freeware resources, e.g. \cite{Bolsig,EEDF}.

While in most swarm studies the very low density of the charged particles justifies neglecting the interaction between them, at higher particle densities, e.g., in plasmas and in particle beams in gases, such interactions may become appreciable. Considering electrons, their interaction in a plasma can be accounted for by adopting a screened Coulomb potential, because of the presence of ions. However, in a beam, where no oppositely charged species are present, the unscreened $1/r$ Coulomb potential has to be considered. The solution schemes of the Boltzmann equation (and also of particle methods) developed to handle electron-electron collisions predominantly focus on plasma conditions, i.e., include screening of the potential. Besides this, they commonly adopt a series of approximations \cite{DD}: (i) search for the distribution function in the form of two terms (``two-term approximation''), (ii) neglect the Coulomb part of the collision integral for the anisotropic part of the distribution function, (iii) treat Coulomb collisions as binary events, and (iv) truncate the range of the electron-electron interaction beyond a characteristic distance. To our best knowledge no efforts have been made recently to avoid these approximations in the solutions of the BE, except of the notable work of Hagelaar \cite{Hagelaar}, who introduced the Coulomb term for the anisotropic part of the distribution function into the two-term solution of the BE. This study has shown that at specific conditions this term may have a significant effect on the results, e.g., on the computed electron mobility. 

It is plausible that the validity of the approximations listed above can only be checked with an approach that is free of these and makes it possible to compute the VDF without any {\it a priori} assumptions. Such a particle-based method relying on first principles has recently been presented in \cite{donko}. This method has been cross-checked with BE solutions for the scenario of the  ``bistabilty'' of the Electron Energy Distribution Function (EEDF) \cite{DD}, an effect that allows the formation of distinctly different EEDF-s at exactly same conditions. While the linearity of the ``pure'' BE excludes the possibility of such multiple solutions, electron-electron collisions change the BE to be nonlinear, which then may have two (or, in principle, multiple) solutions. A generally good agreement was found between the results of the two approaches, with small differences, which were attributed to the approximations adopted in the BE solution scheme, as well as to the (statistical) noise present in the particle simulation \cite{DD}. 

In this paper we focus on two intriguing effects: the {\it Negative Differential Conductivity} (NDC) and the {\it Transient Negative Mobility} (TNM) of electrons in a neutral background gas. In the first case electron-electron (e-e) collisions cause the effect, while in the second case e-e collisions lead to the disappearance of the effect. Our motivation is to confirm the predictions of the approximate BE analysis regarding these effects, via studying them as well by the first-principles particle simulation approach \cite{donko}. 

The physical system considered here is a swarm of electrons in a gas, where no oppositely charged species are present. The motion of the electrons is followed under the influence of a homogeneous electric field, in infinite space. As it will be discussed in more details later (section 3.2), at fixed gas temperature ($T_{\rm g}$) the system is completely characterized by three parameters: (i) the gas number density $N$, (ii) the electron number density, $n_{\rm e}$, and (iii) the electric field, $E$. Alternatively, the parameters $N$, $\eta = n_{\rm e} / N$ (electron to neutral number density ratio), and $E/N$ (reduced electric field) can also be used. 

The physical backgrounds of the two effects investigated are introduced in section 2, while section 3 outlines the computational methods: the particle simulation and the Boltzmann equation approaches. Section 4 presents the results and section 5 gives a short summary.

\section{Effects investigated}

\subsection{Negative Differential Conductivity}

{\it Negative differential conductivity} (NDC) in gases is defined as the decrease of the electron drift velocity with increasing electric field. During the last few decades this phenomenon was comprehensively studied both experimentally and theoretically; reviews of the studies of the NDC effect are given in \cite{p1,p2}. Here we briefly discuss some specific features of the effect, which are important in the context of the present paper.

The conditions under which NDC can be realized were carefully analyzed in \cite{p3,p2} and the occurrence of the effect was explained in terms of the special features of the elastic and inelastic collision cross sections of  electrons with the gas. NDC may occur when an increase in the reduced electric field $E/N$ leads to an abnormally large increase in the elastic collision frequency, which results in a considerable increase in the rate of the randomization of the directions of the electron velocity vectors. In this case the electron drift velocity may decrease even though the mean electron energy grows. According to \cite{p3,p2}, such a situation can be realized in gases (gas mixtures) in which (i) the momentum transfer cross-section, $\sigma_{\rm m}(\varepsilon)$, rises with energy, $\varepsilon $, in a certain energy range and (ii) there exist inelastic energy loss processes in the same range of energy. 

Naturally, whether or not the NDC effect occurs depends on the combination of elastic and inelastic cross sections. The approximate quantitative criterion for the existence of the effect derived in \cite{p3} is written as: \begin{equation}
1 + \frac{{\rm d}\Omega}{{\rm d}\overline{\varepsilon}}<0~~,~~
\Omega(\overline{\varepsilon})=\frac{M}{2m} \varepsilon_{\rm in} 
\frac{\sigma_{\rm in}(\overline{\varepsilon})}{\sigma_{\rm m}(\overline{\varepsilon})} S(\overline{\varepsilon})
\end{equation} where $M$ is the mass of atom (or molecule), $m$ is the mass of electron, $\overline{\varepsilon}$ is the mean electron energy, $\sigma_{\rm in}$ is the cross section for the inelastic process with the threshold energy $\varepsilon_{\rm in}$ and $S$ is a factor, which varies between 0 and 1 for $\overline{\varepsilon} \ll \varepsilon_{\rm in}$  and $\overline{\varepsilon} \gg \varepsilon_{\rm in}$, respectively, and has the effect of ÔsmoothingÕ the rapid jump in $\sigma(\varepsilon_{\rm in})$ in the vicinity of the threshold (see \cite{p3} for more details).

The majority of NDC scenarios was observed either in heavy rare gases with small admixtures of molecular gases, e.g., Ar:CO, Ar:N$_2$,  or in pure molecular gases such as CF$_4$, CH$_4$, CF$_4$ (see \cite{p1,p2} for more details). The momentum transfer cross section of electron collisions with atoms of heavy rare gases and with the molecules listed above (CF$_4$, CH$_4$, CF$_4$) possesses a Ramsauer-Townsend (R-T) minimum and grows sharply with energy above this minimum. The second condition, the presence of inelastic energy losses, is provided by the excitation of vibrational levels of molecules. 
It should be noted that, according to the calculations presented in Refs. \cite{ndc-new,add2,add3}, the NDC effect can also be induced by electron attachment and/or electron impact ionization processes.

Though inelastic or non-conservative collisions are generally needed for the appearance of the NDC, there are at least three exceptions from this rule. 

The first is that NDC was predicted in Xe:He and Kr:He mixtures \cite{p4,p5}. The qualitative explanation of the effect was as follows: as He atoms are light, electrons lose rather noticeable portion of energy in elastic collisions with He atoms. This way elastic collisions with He atoms in Xe:He and Kr:He mixtures play a role similar to inelastic collisions in rare gas-molecular gas mixtures. The quantitative criterion for the NDC effect in mixtures of rare gases was derived in \cite{ndc-new}. In \cite{p6} the observation of the effect was reported in Xe:He mixtures.

The second case is the NDC in dense gases or liquids. In this case the effect arises purely as a consequence of the coherent scattering of electrons from a structured material (see \cite{WR} and references therein).

The third case is that NDC was predicted in pure heavy rare gases at conditions when the electron to neutral number density is relatively high \cite{p7}, e.g., in Xe at $\eta = n_{\rm e}/N > 4 \times 10^{-9}$. While the electron drift velocity in heavy rare gases is ``normally'' a monotonic increasing function of the reduced electric field, electron-electron collisions have been found to change this behavior and to result in the appearance of the NDC effect. The origin of the effect has been attributed to specific changes in the shape of the electron energy distribution function (EEDF) caused by e-e collisions (see \cite{p1} and \cite{p7} for more details). The possibility of the experimental verification of the NDC phenomenon in Xe was analyzed in \cite{p8}. However, so far there were no experimental observations of the effect in pure heavy rare gases and the theoretical predictions were based only on Boltzmann equation analysis that includes the approximations listed in section 1). It is also worth noting that in discharge plasmas, where a high concentration of excited and charged particles is present, superelastic collisions of electrons with excited atoms and molecules, as well as Coulomb collisions may have a significant impact on the NDC effect (see comments in \cite{p1}).

\subsection{Transient Negative Mobility}

{\it Transient Negative Mobility} (TNM) is a scenario when the temporal change of the EEDF shape is much faster than the variation of the electron number density and the electron mobility becomes negative during the relaxation of the EEDF.

The conditions necessary for the electron mobility, $\mu_{\rm e}$, to be negative -- discussed in several papers -- are derived from the two equivalent expressions: 
\begin{subequations}
\begin{eqnarray}
\mu_{\rm e}= -\frac{e}{3N}\sqrt{\frac{2}{m}}\int\displaylimits_0^\infty \frac{\varepsilon}{\sigma_{\rm m}(\varepsilon)}
\frac{{\rm d}f_0(\varepsilon)}{{\rm d} \varepsilon} {\rm d} \varepsilon = \label{eq:mu1} \\ 
\frac{e}{3N}\sqrt{\frac{2}{m}}\int\displaylimits_0^\infty f_0(\varepsilon) \frac{{\rm d}}{{\rm d} \varepsilon} \biggl[ \frac{\varepsilon}{\sigma_{\rm m}(\varepsilon)} \biggr]
 {\rm d}\varepsilon,
\label{eq:mu2}
\end{eqnarray}
\end{subequations}
where $e$ and $m$ are the charge (absolute value) and the mass of an electron and $f_0(\varepsilon)$ is isotropic part of the EEDF, which is normalized as 
\begin{equation}
\int\displaylimits_0^\infty \sqrt{\varepsilon} f_0(\varepsilon) {\rm d} \varepsilon = 1. \nonumber
\end{equation}
We note that (\ref{eq:mu2}) has been derived by integrating (\ref{eq:mu1}) by parts.

The first condition for the appearance of the negative mobility (see (\ref{eq:mu1})) is  ${\rm d}f_0/{\rm d} \varepsilon > 0$ in a certain energy range, implying that $f_0(\varepsilon)$ should have a local maximum at a particular (nonzero) energy. The distribution function with such properties is called to have an ``inverse shape''. The second condition (see (\ref{eq:mu2})) is that the inequality ${\rm d}(\varepsilon/\sigma_{\rm m}(\varepsilon))/{\rm d} \varepsilon < 0$ should be satisfied, which means that $\sigma_{\rm m}(\varepsilon)$ should increase faster than linearly with energy. Note that both conditions are necessary but not sufficient for the electron mobility to be negative.

To understand the physical nature of the appearance of negative mobility let us suppose that $f_0(\varepsilon)$ is a delta function located at an energy, where the condition \\ ${\rm d}(\varepsilon/\sigma_{\rm m}(\varepsilon)) / {\rm d} \varepsilon < 0$ is met. Electrons in the gas can be divided into two groups: (i) moving against and (ii) along the direction of the electric field. The first group gains energy from the electric field. The elastic collision frequency for these electrons increases (since $\sigma_{\rm m}(\varepsilon)$ sharply grows with energy), and the velocity direction is quickly randomized, which leads to decrease in the mean velocity against the electric field. On the contrary, electrons in the second group lose their energy, the elastic collision frequency for these electrons decreases, and the mean velocity along the electric field grows. As a result, a situation may be realized when the average electron velocity (drift velocity) is along the electric field, i.e. the electron mobility is negative.

The required, faster-than-linear increase of $\sigma_{\rm m}(\varepsilon)$ takes place for heavy rare gases in the energy range above the R-T minimum. For this reason, heavy rare gases were considered to be the main components of the gas mixtures in all papers devoted to studies of negative electron mobility. The distribution function with ``inverse shape'' (as defined above) can form and thus negative electron mobility can appear under various physical conditions: in plasmas of heavy rare gases during EEDF relaxation (TNM), in steady or decaying plasmas of heavy rare gases with electronegative admixture, as well as in optically excited plasmas of heavy rare gases with admixtures of metal atoms (see the reviews \cite{p1,p9} and references therein). 

The TNM phenomenon was mentioned first in Refs. \cite{p10,p11}, where the time evolution of the EEDF in an electric field in Ar, Kr and Xe was studied theoretically. Assuming an initial delta function shape for $f_0(\varepsilon)$  located at a given energy $\epsilon_0$, it was found that at the electron mobility in xenon is negative within a short time interval during the EEDF relaxation. Later, TNM was predicted in calculations of the EEDF time-variation in Ar after switching off the external electric field \cite{p12}. There, the initial $f_0(\varepsilon)$ was taken to be equal to the steady state EEDF in the initial electric field, i.e., it was not of an ``inverse shape''. In this case, an EEDF of inverse shape was formed during the course of the relaxation process; the TNM effect was found to appear at $E/N \geq$ 0.15 Td. In \cite{p12} no calculations were performed for Kr and Xe, but it was stated that the TNM effect should take place in these gases, too. Such calculations for Xe were carried out in \cite{p9,p13}, where the reduced electric field was switched from a relatively high (1--5 Td) to a low (0.01 Td) value. It was shown that during the EEDF relaxation there exists a time interval where the electron mobility is negative. The TNM effect was also numerically studied in \cite{p14} (for pure Xe) and \cite{p15} (for Xe:Cs mixture), in which the initial $f_0(\varepsilon)$ was assumed to be a narrow Gaussian distribution around a certain energy. 

In an experimental demonstration of the effect \cite{p16}, the displacement current resulting from the motion of electrons and ions following the ionization of the gas  (Xe, 20 atm) between two parallel plate electrodes by a short (10 ns) X-ray pulse was measured. At a low electric field of 1.16$\times 10^{-3}$ Td applied across the electrodes the measured current was found to be negative during $\sim$10 ns after the end of the ionizing pulse. Since electrons are the main charge carriers in the ionized gas and the electron removal processes were negligible within the time interval considered ($\sim$100 ns) the measured current was assumed to be proportional to the transient  negative electron mobility \cite{p17}. 

In most of the theoretical papers devoted to the TNM effect the electron concentration was assumed to be small and e-e collisions were not taken into account in the calculations. However, the inverse-shaped distribution function needed for the appearance of the effect differs fundamentally from the Maxwellian function, and one expects that e-e collisions shall have appreciable influence on the EEDF even at low ionization degrees. The influence of e-e collisions on the EEDF shape and the value of transient electron mobility was studied in \cite{p13} for the case of Xe. It was shown that the TNM effect disappears at $n_{\rm e}/N \geq 10^{-8}$. Let us note that this theoretical prediction was based on the Boltzmann equation analysis (that includes the approximations listed in section 1) and the possible effects of the approximations involved remained unchecked.

\section{Computational methods}

Below we give a brief description of the methods used here: the particle simulation scheme is presented in section 3.1, while the solution of the Boltzmann equation is discussed in section 3.2. A more detailed description of the two computational approaches can be found in \cite{DD}. 

\subsection{Particle simulation}

The simulation scheme is based on a combination of a Molecular Dynamics (MD) technique and a Monte Carlo (MC) approach \cite{donko}. The MD part describes the many-body interactions driven by the inter-particle Coulomb potential within the classical electron gas, while the MC part handles the interaction of the electron gas with the background (atomic) gas. We consider a fixed number of electrons, (at the relatively low reduced electric fields adopted here ionization is very unlikely). The equation of motion of the $i$-th electron is:
\begin{equation}
m \frac{{\rm d}^2 {\bf r}_i}{{\rm d}t^2} = \sum_{i \neq j} {\bf F}_{ij} - e {\bf E}.
\end{equation}
The sum on the right hand side represents the force exerted on particle $i$ by all other ($j \neq i$) particles {\it and their periodic images} located in spatial replicas of the simulation box. These images have to be included in the proper determination of the interparticle forces in the case of the un-truncated, infinite-range Coulomb potential. Note that for our conditions no screening of the potential takes place due to the absence of oppositely charged species. This summation is a key issue and needs a special approach; our choice is the Particle-Particle Particle-Mesh (PPPM) algorithm, described in details in \cite{HE}. The second term on the right hand side of the above equation is a contribution due to the external electric field.   

The equations of motion of the electrons are numerically integrated with a discrete time step $\Delta t$, for which the (upper) limit is defined by the stability of the integration of the equations of motion in the case of the closest approach of two electrons, $r_{\rm min} = e^2/(4 \pi \epsilon_0 \varepsilon_{\rm max})$. Here $\varepsilon_{\rm max}$ is a pre-defined maximum energy \cite{HE}, which has to be chosen carefully, to ensure that the probability of finding electrons with $\varepsilon > \varepsilon_{\rm max}$ is vanishingly small at the conditions considered. 

The electron gas and the background gas interact via e$^-$+Xe collisions. The probability of an e$^-$+Xe collision during a time step $\Delta t$ is calculated as:
\begin{equation}
P_{\rm coll} = 1 - {\rm exp}[-n \sigma_{\rm tot}(g) g \Delta t],
\end{equation}
where $\sigma_{\rm tot}$ is the total scattering cross section, $g = |{\bf g}|$, with ${\bf g}= {\bf v} - {\bf V}$ being the relative velocity between the electron and a Xe atom with a velocity ${\bf V}$ randomly chosen from a Maxwellian background of gas atoms having a temperature $T_{\rm g}$. This probability is calculated for each electron in each time step, and decision about the occurrence of a collision is made by comparing $P_{\rm coll}$ with a random number. Another random number is used to select that actual process to be executed, based on the values of the respective cross sections of the individual processes at the actual value of the relative energy of the collision partners \cite{MC}. Collisions are executed in the center-of-mass frame, and are considered to be isotropic. 

The simulations starts from a random particle configuration, the equilibration of the system is monitored by observing the velocity moments of the VDF as a function of time. Measurements in the steady state are started when these moments have acquired stable values \cite{donko}.

\subsection{Solution of the Boltzmann equation}

The spatially homogeneous Boltzmann equation for electrons is:
\begin{equation}
\frac{\partial f({\bf v})}{\partial t} - \frac{e {\bf E}}{m} ~ \nabla_{\bf v} f({\bf v}) =C,
\label{eq:be1}
\end{equation}
where $C$ is the collision integral, which, in the present study, accounts for the following processes: elastic scattering of electrons from atoms (the corresponding part of the collision integral is designated as $C_{\rm m}$), excitation of electronic states and ionization of atoms by electron impact from the ground state ($C_{\rm in}$), as well as electron-electron collisions ($C_{\rm e}$): $C = C_{\rm m} + C_{\rm in} + C_{\rm e}$. 

The conventional method of solving eq. (\ref {eq:be1}) is based on the expansion of the distribution function $f({\bf v})$ in Legendre polynomials, $P_n(\cos \Theta)$. Retaining the two first terms only we arrive at the two-term approximation:
\begin{equation}
f({\bf v})= f_0(v) + f_1(v) \cos \Theta, 
\label{eq:be2} 						
\end{equation}
where $v$ is the magnitude of the velocity, $\Theta$ is the angle between ${\bf v}$ and $-{\bf E}$, $f_0(v)$ is the symmetrical (isotropic) part of the distribution function and $f_1(v)$ (the anisotropic part) describes the directed motion of the electrons along the electric field. The substitution of expansion (\ref{eq:be2}) into equation (\ref{eq:be1}) leads to:
\begin{equation}
\frac{\partial f_0}{\partial t} - \frac{e E}{3 m v^2} \frac{\partial}{\partial v} (v^2 f_1) = C_{\rm 0,m} + C_{\rm 0,in} + C_{\rm 0,e} 
\label{eq:f0}
\end{equation}
and
\begin{equation}
\frac{\partial f_1}{\partial t}- \frac{e E}{m} \frac{\partial f_0}{\partial v} = C_{\rm 1,m} + C_{\rm 1,in} + C_{\rm 1,e}. 
\label{eq:f1}
\end{equation}

The collision integrals $C_{\rm 0,m}$ and $C_{\rm 1,m}$ can be written as \cite{int2}:
\begin{eqnarray}
C_{\rm 0,m} =  \frac{1}{2 v^2} \frac{\partial}{\partial v} \Biggl[ \frac{2m}{M} \nu_{\rm m} v^2 \biggl( \frac{k_{\rm B}T_{\rm g}}{m} \frac{\partial f_0}{\partial v} + v f_0 \biggr) \Biggr], \\
C_{\rm 1,in} = -\nu_{\rm m} f_1, 
\end{eqnarray}
where $\nu_{\rm m} = N \sigma_{\rm m} v$ is the momentum transfer frequency and $\delta = 2m/M$ is the average fraction of the energy lost by the electrons in one elastic collision with atom ($M$ is the mass of the gas atom). The rate of the electron energy loss due to elastic collisions is characterized by the frequency $\nu_{\rm u} = \delta \nu_{\rm m}$.

In the present work the ionization of atoms by electron impact is treated as a conservative process, i.e., as an excitation of an electronic state, of which the energy is equal to the ionization potential. All inelastic processes are supposed to result in isotropic scattering. With these assumptions the collision integrals $C_{\rm 0,in}$ and $C_{\rm 1,in}$ can be written as \cite{SJB,newrefx}:
\begin{eqnarray}
C_{\rm 0,in} = \frac{1}{v^2} \sum_k \Bigl[ f_0(v_k) v_k^2 \nu_{\rm in}^{[k]}(v_k) -f_0(v) v^2 \nu_{\rm in}^{[k]}(v) \Bigr],   \\
C_{\rm 1,in} = -\sum_k \nu_{\rm in}^{[k]} f_1, 
\end{eqnarray}
where $\nu_{\rm in}^{[k]} = N \sigma_{\rm in}^{[k]} v$ is the frequency of the excitation of $k$-th electronic state ($\sigma_{\rm in}^{[k]}$ is the corresponding cross section), $\varepsilon_k$ is the energy of $k$-th electronic state, and $v_k = \sqrt{v^2 + 2 \varepsilon_k / m}$. 

It is known that in the case of Coulomb collisions the calculation of the pair-collision frequency encounters a distinctive problem, namely, the logarithmic divergence of the frequency at small scattering angles. To overcome this problem it is assumed that the Coulomb potential acts only up to a certain finite distance $r_{\rm max}$. Then the expression for the term $S_{\rm 0,e}$ is written as follows \cite{int2,SJB}:
\begin{eqnarray}
&& S_{\rm 0,e} = \frac{1}{v^2} \frac{\partial}{\partial v}\Biggl\{ v^2 \nu_{\rm e} \biggl[ A_1(f_0) v f_0 + A_2 (f_0) \frac{\partial f_0}{\partial v} \biggr]  \Biggr\}, \\
&& A_1(f_0) = 4 \pi \int_0^v (v')^2 f_0(v') {\rm d}v',\\
&& A_2(f_0) = \frac{4 \pi}{3} \Biggl[ \int_0^v (v')^4 f_0(v') {\rm d}v' +  \nonumber \\ &&~~~~~~~~~~~~~~~v^3 \int_v^\infty v' f_0(v') {\rm d}v' \Biggr],
\end{eqnarray}
where $f_0(v)$ is normalized as:
\begin{equation}
4 \pi \int_0^\infty v^2 f_0(v) {\rm d}v = 1 \nonumber
\end{equation}
and 
\begin{eqnarray}
\nu_{\rm e} = 2 \pi n_e \biggl( \frac{e^2}{4 \pi \epsilon_0 m}\biggr)^2 \frac{1}{v^3} \ln \Biggl[ 1 + \biggl(\frac{r_{\rm max}}{r_0}\biggr)^2 \Biggr]~~, \label{eq:nue} \\ \nonumber
r_0 = \frac{e^2}{4 \pi \epsilon_0 m v^2}. 
\end{eqnarray}

The value of $r_0$ is usually estimated as $r_0 = e^2 / 4 \pi \epsilon_0 2 \overline{\varepsilon}$ in the calculations, where $\overline{\varepsilon}$ is the mean electron energy. As to the cutoff distance, for the case of plasmas it is generally assumed that $r_{\rm max}$ is equal to the Debye length. For the case of swarm conditions considered here we use the approximation that the cutoff distance is equal to the half of the average distance between the electrons:
\begin{equation}
r_{\rm max} = 0.5 n_{\rm e}^{-1/3}.
\label{eq:rmax}
\end{equation}
The physical ground of this approximation is as follows \cite{DD}. If the impact parameter (of test electron relative to a given electron) is higher than the half the average distance between electrons in the gas, then the influence of this electron on the test one becomes weaker than the influence of the neighboring one. Note that in an ideal electron gas $r_{\rm max}/r_0 = \overline{\varepsilon} / (e^2 / 4 \pi \epsilon_0 n_{\rm e}^{-3}) \gg 1$ and the ``1'' in the expression under the logarithm in (\ref{eq:nue}) can be omitted. 

The term $S_{\rm 1,e}$ in eq. (4), it is very complex (see comments in \cite{int2}) and, as a rule, it is neglected assuming that $\nu_{\rm e} \ll \nu_{\rm m}$. To our best knowledge, the influence of this term on the EEDF shape and plasma characteristics was analyzed in a recent paper only \cite{Hagelaar}. In the present work the term $S_{\rm 1,e}$ is neglected. Moreover, if the characteristic time of the changes of the plasma parameters is essentially greater than $\nu_{\rm m}^{-1}$, the time derivative in eq. (\ref{eq:f1}) can be omitted, resulting in:  
\begin{equation}
f_1  = \frac{e E}{m (\nu_{\rm m} +\sum_k \nu_{\rm in}^{[k]})} \frac{\partial f_0}{\partial v},
\label{eq:f1final}
\end{equation}
while for $f_0$ we arrive at
\begin{eqnarray}
\frac{\partial f_0}{\partial t} - \frac{e E}{3 m v^2} \frac{\partial}{\partial v} \Biggl( v^2  \frac{e E}{m (\nu_{\rm m} +\sum_k \nu_{\rm in}^{[k]})} \frac{\partial f_0}{\partial v} \Biggr) = \nonumber \\ C_{\rm 0,m} +  C_{\rm 0,in} + C_{\rm 0,e}.
\label{eq:be17}
\end{eqnarray}

It follows from (\ref{eq:be17}) that, at fixed gas temperature and in the absence of e-e collisions, the parameter for the steady state solution of eq. (\ref{eq:be17}) is the reduced electric field, $E/N$. If the e-e collisions are taken into account, there are three parameters: $E/N$, $n_{\rm e}/N$ and $n_e$ (or, equivalently, $E$, $N$, and $n_{\rm e}$). The electron number density is an independent parameter since the logarithmic term in eq. (\ref{eq:nue}) (in the Coulomb logarithm) depends on $n_{\rm e}$. Actually, at fixed $E/N$ and $n_{\rm e}/N$ values the dependence of $f_0$ on $n_e$ is rather weak, since $n_e$ is under the logarithm.

To obtain its numerical solution, eq. (\ref{eq:f1final}) it is rewritten with energy as a variable. The steady-state equation is solved by an iteration method similar to that described in \cite{r6,r7}. In the case of calculation of the time-dependent solution of the BE the time step should be as small as $\Delta t \ll \nu_{\rm u}^{-1}$ and $\Delta t \ll \nu_{\rm e}^{-1}$, where $\nu_{\rm u}$ is the frequency, which characterizes the rate of the electron energy loss due to elastic and inelastic collisions.

\section{Results}

Below we present our computational results for the two effects investigated. The results for the Negative Differential Conductivity are discussed in section 4.1, while the Negative Transient Mobility occurring during the relaxation of the electron swarm, induced by a sudden change of the external electric field, is analyzed in section 4.2. All calculations adopt the cross sections given in \cite{momcs,Hayashi}.

\subsection{Negative Differential Conductivity}

\begin{figure}
\resizebox{0.44\textwidth}{!}{
  \includegraphics{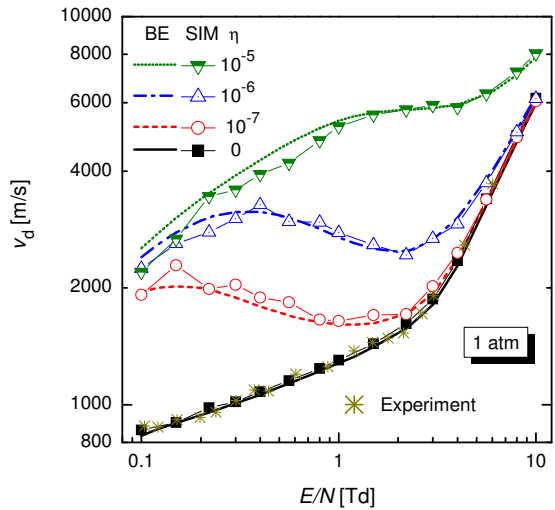}}      
\caption{(color online) Drift velocity of the electrons in Xe as a function of the reduced electric field, at $p=$ 1 atm and $T_{\rm g}$ = 300 K, for different electron to neutral density ratios ($\eta$). The source of experimental data is \cite{vdexp}. BE: solutions of the Boltzmann equation, SIM: results of the particle simulation.}
\label{fig:f1}       
\end{figure}

The swarm characteristics are studied at gas pressures of $p$ = 1 atm and 10 atm, at a fixed temperature of $T_{\rm g}$ = 300~K (that defines the gas number density $N$), for different values of the electron to gas number density ratio $\eta$. Figure \ref{fig:f1} shows the electron drift velocity at $p$ = 1 atm, as a function of the reduced electric field, $E/N$, between 0.1 Td and 10 Td. The particle simulations have been carried out with 500 electrons. At 1 atm pressure the data points originate from runs comprising $8 \times 10^8$ measurement time steps (which follow an initialization period during which the stationary state is established). To define the simulation time step we adopt a maximum energy (see section 3.1) of $\varepsilon_{\rm max} \sim 10$ eV, which leads to $\Delta t \sim 10^{-16}$ s. Such a short time step results in very demanding computations, the computational speed is typically $\sim$ 1 ns / day, on a single CPU. The above number of time steps corresponds to about 30 days of runtime on a single CPU for each data point. In the calculations at 10 atm pressure (see below) $4 \times 10^8$ measurement time steps were executed. 

\begin{figure}
\resizebox{0.44\textwidth}{!}{
  \includegraphics{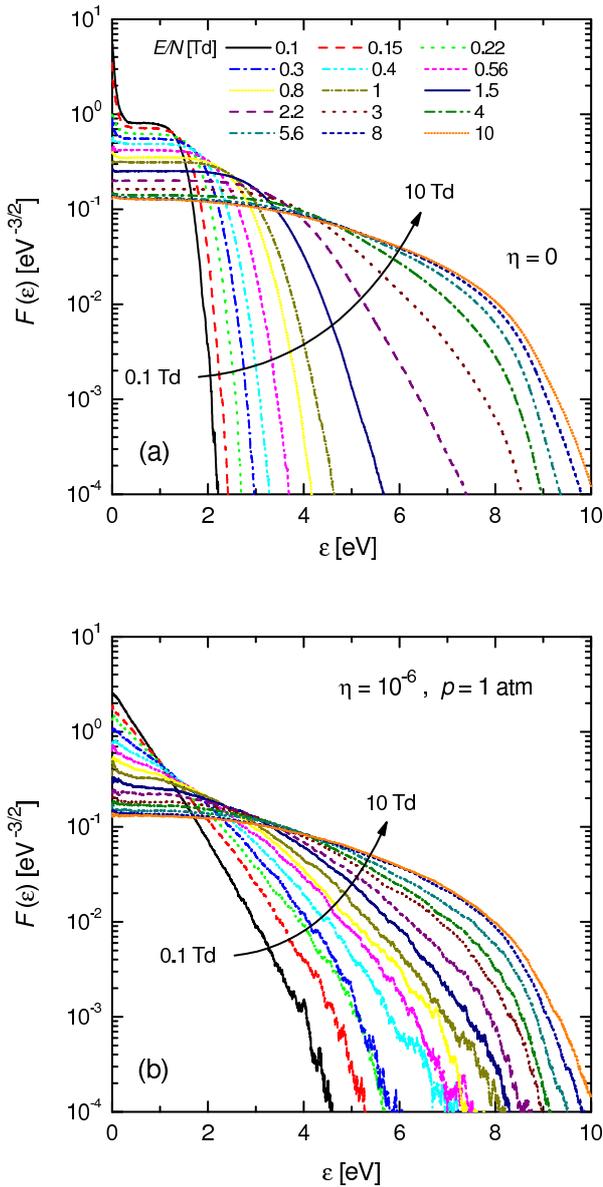}}      
\caption{(color online) EEDF-s in Xe (results of the particle simulation) as a function of the reduced electric field, for (a) $\eta =0$ and (a) $\eta =10^{-6}$, at $p=$ 1 atm. The legend in (a) holds for both panels.}
\label{fig:f2}       
\end{figure}

In the absence of electron-electron collisions the drift velocity $v_{\rm d}$ is a monotonically increasing function of $E/N$, the particle simulation and BE results agree well with each other, as well as with the experimental results of Ref. \cite{vdexp}. With e-e interactions included, our results cover electron to gas density ratios $10^{-7} \leq \eta \leq 10^{-5}$. A pronounced negative slope of the $v_{\rm d}(E/N)$ curve -- indicating NDC -- shows up at density ratios of $\eta = 10^{-7}$ and $10^{-6}$, while the NDC effect disappears at the highest electron density of $\eta = 10^{-5}$. The results obtained by our two methods are in a good agreement, no systematic deviations can be found, considering the scattering of the data points (due to limitations of the statistics) obtained from the particle simulation. This confirms that for these conditions that approximations adopted in the BE solution are justified.

It is worth mentioning that the strongest effect of the e-e interactions appears at low $E/N$, while at 10 Td the drift velocity is insensitive on $\eta$ unless it reaches a high value. A similar behavior is found for the EEDF-s computed with the particle simulation method for the different conditions (for $p$ = 1 atm). Figure \ref{fig:f2}(a) shows a series of EEDF-s obtained for $\eta =0$, for different $E/N$ values, while Fig. \ref{fig:f2}(b) shows a similar set of EEDF-s obtained at $\eta = 10^{-6}$. A pronounced effect of the e-e interactions -- Maxwellization of the EEDF -- is well visible at the low $E/N$ conditions in Fig. \ref{fig:f2}(b). With increasing electric field this effect diminishes as the electron energy gain from the electric field starts to be balanced by inelastic collisions; at $E/N$ = 10 Td, e.g., the change of the EEDF due to e-e collisions is negligible at $\eta = 10^{-6}$, cf. Figs. \ref{fig:f2}(a) and (b).

\begin{figure}
\resizebox{0.44\textwidth}{!}{
  \includegraphics{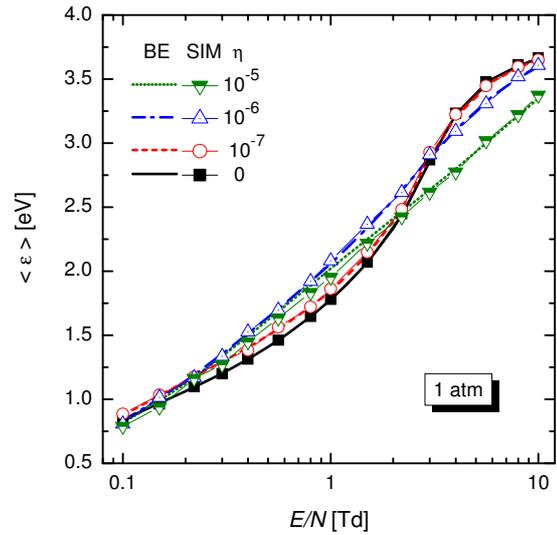}}      
\caption{(color online) Mean energy of the electrons as a function of the reduced electric field, at $p=$ 1 atm and different electron to neutral density ratios ($\eta$). BE: solutions of the Boltzmann equation, SIM: results of the particle simulation.}
\label{fig:f3}       
\end{figure}

In contrast with the pronounced effect of e-e interaction on the drift velocity, the mean electron energy, $\langle \varepsilon \rangle$ is less influenced by this effect. As Fig. \ref{fig:f3} displays, the increase of $\langle \varepsilon \rangle$ is monotonic for all values of $\eta$ covered here (recall the discussion of the physical basis of the effect in section 2.1). It is worth noting, however, that at low $E/N$ values the e-e collisions increase the mean electron energy, while an opposite effect is found above $E/N \sim 2-3$ Td. 

The e-e interactions always lead to a distortion of the EEDF towards Maxwellian, depending on the shape of the EEDF, however, this tendency may have opposite consequences: (i) the formation of a high energy tail of the EEDF, resulting from this Maxwellization may increase $\langle \varepsilon \rangle$ if the EEDF is initially confined at low energies; (ii) whenever the EEDF has a high population at medium energies (at several eV-s), this population may decrease due to the Maxwellization, and consequently, $\langle \varepsilon \rangle$ may decrease. The increase of $\langle \varepsilon \rangle$ at low $E/N$ can be explained by the first effect (see Fig.~\ref{fig:f2}(b)). At high $E/N$, e.g., at 10 Td, however, the second scenario takes place, the Maxwellization depopulates the EEDF in the range of medium energies (4--9 eV) thereby decreasing the mean energy.

\begin{figure}
\resizebox{0.44\textwidth}{!}{
  \includegraphics{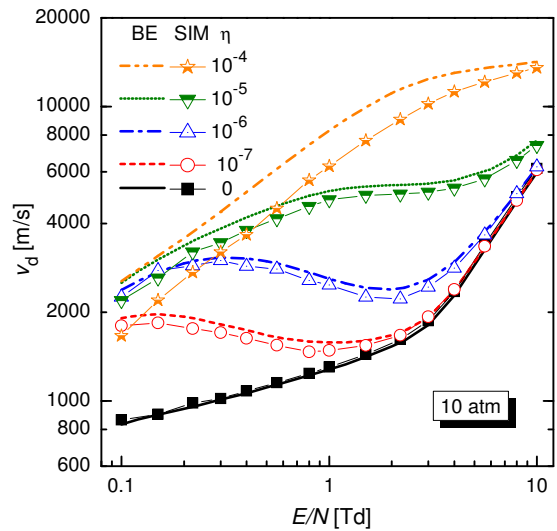}}      
\caption{(color online) Drift velocity of the electrons in Xe as a function of the reduced electric field, at $p=$ 10 atm and different electron to neutral density ratios ($\eta$). BE: solutions of the Boltzmann equation, SIM: results of the particle simulation.}
\label{fig:f4}       
\end{figure}

The effect of e-e interactions on the drift velocity at the higher pressure of $p$ = 10 atm (and $T_{\rm g}$ = 300 K) is demonstrated in Fig.~\ref{fig:f4}. This figure also shows data obtained at $\eta$ = 10$^{-4}$. At this high value of the density ratio, as well as at $\eta$ = 10$^{-5}$ the drift velocity is a monotonic function of $E/N$, while at $\eta$ = 10$^{-6}$ and $\eta$ = 10$^{-7}$ the NDC effect is clearly observed. The data obtained from the BE solution and the particle simulations show the same trend, but differ systematically: the particle simulations yield lower values of $v_{\rm d}$. The reason of this discrepancy at these conditions -- considering the low noise of the simulation data -- should be searched for in the approximations adopted in the solution scheme of the BE. 

\begin{figure}
\resizebox{0.4\textwidth}{!}{
  \includegraphics{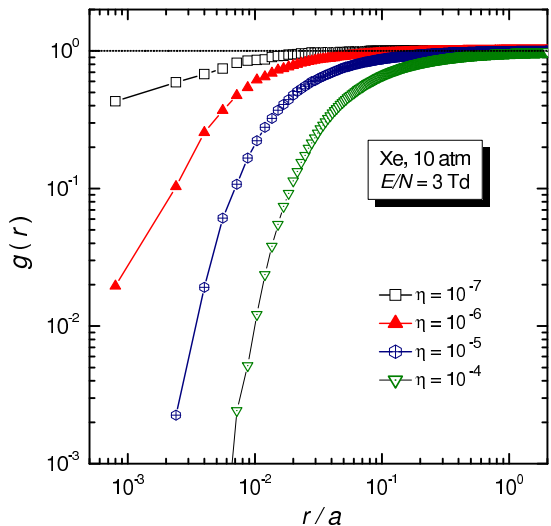}}      
\caption{(color online) Pair distribution functions of electrons in Xe at $p=$ 10 atm, at $E/N$ = 3 Td and different electron to neutral density ratios ($\eta$). The distance $r$ is normalized by the Wigner-Seitz radius $a = (4 \pi n_{\rm e} / 3)^{-1/3}$. The dotted horizontal line at $g(r)=1$ corresponds to an ideal gas.}
\label{fig:f5}       
\end{figure}

\begin{figure}
\resizebox{0.44\textwidth}{!}{
  \includegraphics{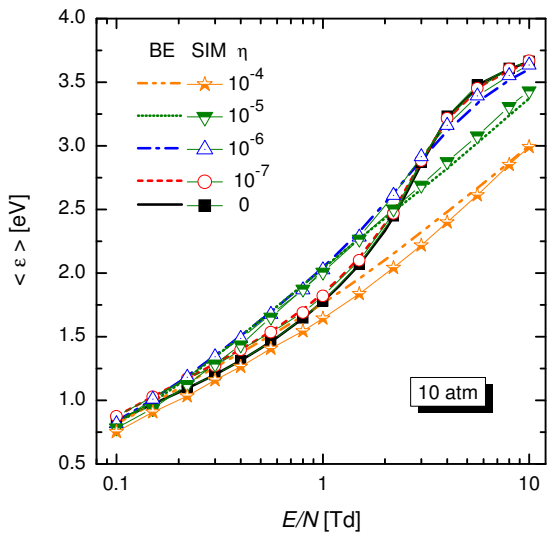}}      
\caption{(color online) Mean energy of the electrons in Xe as a function of the reduced electric field, at $p=$ 10 atm and different electron to neutral density ratios ($\eta$). BE: solutions of the Boltzmann equation, SIM: results of the particle simulation.}
\label{fig:f6}       
\end{figure}

It is noted that at such a high pressure the electron density is rather high as well, e.g., $n_{\rm e} = 2.45 \times 10^{21}$ m$^{-3}$, at $\eta$ = 10$^{-5}$. At such densities the electron gas becomes increasingly non-ideal, as indicated by the development of correlations between the positions of the electrons, which can be examined (quantified) by the pair distribution function, $g(r)$, that gives the density distribution of particles around a test particle, compared to a uniform distribution, characteristic for an ideal gas. Figure~\ref{fig:f5} shows the $g(r)$ functions computed in the particle simulations for different $\eta$ values, at $p=$ 10 atm and $E/N$ = 3 Td. The development of a ``correlation hole'' at low particle separations, due to the increasing role of the inter-particle potential energy is clearly seen with increasing $\eta$. The non-ideal nature of the electron gas is not captured in the BE solution scheme, so this effect might contribute to the differences seen in the drift velocity data, although the electron gas is only slightly non-ideal even at $\eta = 10^{-4}$, where the mean potential energy / mean kinetic energy ratio is in the order of 0.02.

The behavior of the mean electron energy, $\langle \varepsilon \rangle$, as a function of $E/N$ and $\eta$ is shown in Fig.~\ref{fig:f6} for $p$ = 10 atm. At high electric fields the decrease of the mean energy -- caused by the depopulation of the EEDF at medium energies -- is more pronounced as compared to the case of $p$ = 1 atm (cf. Fig.~\ref{fig:f3}). At this high pressure the dependence of $\langle \varepsilon \rangle$ on $\eta$ is clearly non-monotonic even at low $E/N$.

\subsection{Transient Negative Mobility}

The Transient Negative Mobility effect in Xe is investigated at $p=10$ atm and $T_{\rm g}=$ 300 K. Using both methods convergent solutions for a stationary state at $E/N$ = 2.2 Td were first found. This state was used as the initial state for the time-dependent computations following a change of $E/N$ to 0.01 Td, at $t=0$. We have computed the swarm characteristics for two electron to gas density ratios: $\eta = 0$ and $\eta = 10^{-7}$. 

\begin{figure}
\resizebox{0.42\textwidth}{!}{
  \includegraphics{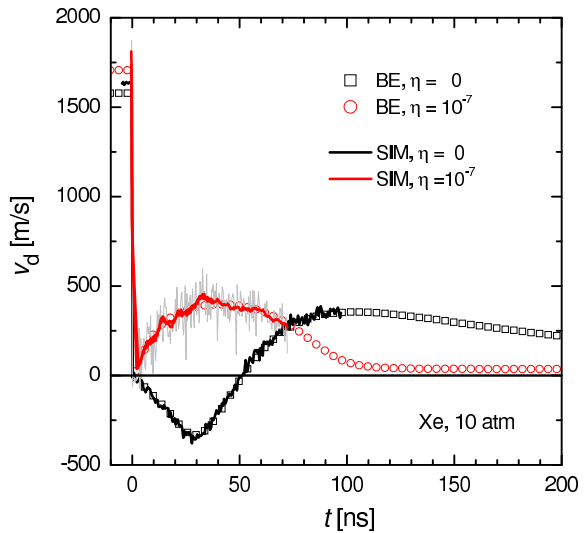}}      
\caption{(color online) Time dependence of the electron drift velocity in Xe, following a change of the electric field at $t=0$, from $E/N$ = 2.2 Td to 0.01 Td, at $p$ = 10 atm, in Xe ($T_{\rm g}$=300 K). In the case of the particle simulations for $\eta = 10^{-7}$ the raw results (average of 30 simulation runs) are shown by the light grey line, the thick red line is a smoothed curve. Symbols represent the results of the BE solutions.}
\label{fig:f7}       
\end{figure}

\begin{figure}
\resizebox{0.42\textwidth}{!}{
  \includegraphics{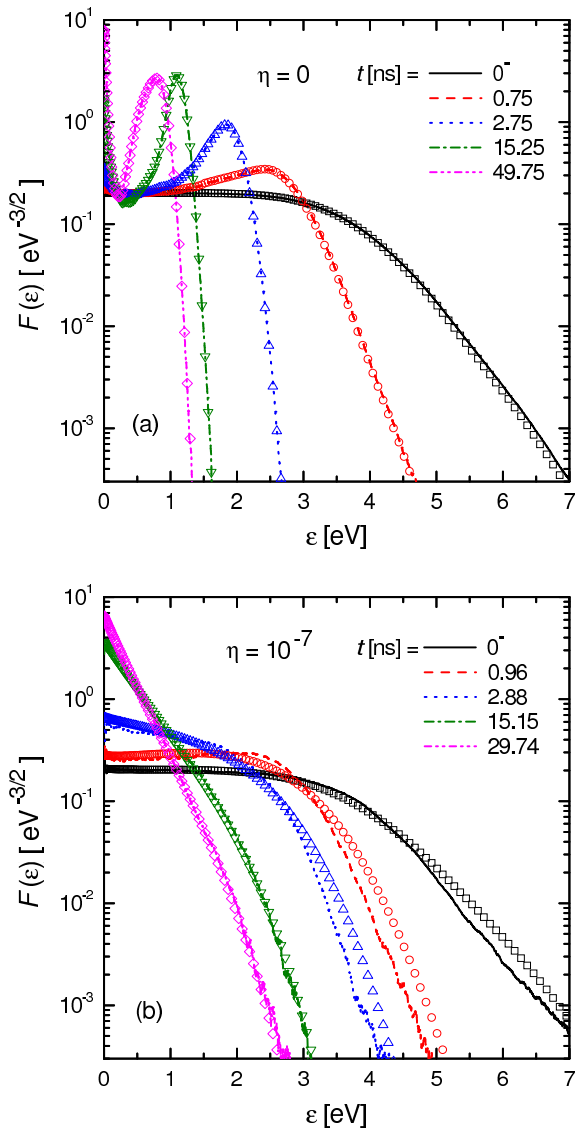}}      
\caption{(color online) Temporal relaxation of the EEDF following the change $E/N$ = 2.2 Td $\rightarrow$ 0.01 Td at $t=0$, for $\eta=0$ (a) and $\eta = 10^{-7}$ (b), as derived from the particle simulation (lines) and from the solution of the Boltzmann equation (symbols). $t= 0^-$ corresponds to the initial steady state at $E/N$ = 2.2 Td. At $t>0$ the curves show averages of the EEDF-s over 0.5 ns intervals for $\eta=0$ and over 0.384 ns intervals for $\eta=10^{-7}$; $t$ is the center of the averaging intervals. ($p$ = 10 atm.)}
\label{fig:f8}       
\end{figure}

The time-dependence of the electron drift velocity is displayed in Fig.~\ref{fig:f7}, for both of these conditions, as obtained from the two computational methods. In the case of particle simulation the data at $\eta=0$ originate from averaging the results of 10 independent simulations each comprising 10$^5$ particles. At $\eta = 10^{-7}$ we used 30 independent simulations, each comprising 1000 electrons. The runtime of each of these simulations was about 100 days, the data obtained this way cover the $0 \leq t \leq 100$ ns range at $\eta = 0$ and the $0 \leq t \leq 80$ ns range at $\eta = 10^{-7}$, while the data originating from the BE solution extend to several hundred nanoseconds. In the time-dependent solutions of the Boltzmann equation the time step was set to $\Delta t = 10^{-13}$ s.  

The results show a very good agreement for both cases studied over the whole time domain for which the particle simulation data are available.  The development of the negative electron mobility at $\eta = 0$ within the time domain $t < $ 50 ns is clearly seen. A drift velocity of about $v_{\rm d} \approx -350$ m/s is reached at $t = 30$ ns. Beyond $t=$ 50 ns the drift velocity becomes positive, following a broad peak at 100 ns it slowly relaxes. This relaxation to the stationary value at $E/N$ = 0.01 Td takes about 1000 ns. In the presence of e-e interactions a completely different behavior of $v_{\rm d}(t)$ is found. Following the decay of the electric field, the drift velocity rapidly drops to almost zero and then exhibits a broad positive peak during which $v_{\rm d}$ reaches $\approx$ 400 m/s, which is an order of magnitude higher than the stationary velocity reached after $t \sim$ 120 ns, where the swarm is relaxed at the ``new'' conditions. 

The temporal relaxation of the EEDF is plotted in Figs.~\ref{fig:f8}(a) and (b), for $\eta=0$ and $\eta = 10^{-7}$, respectively. In order to achieve an acceptable statistics the particle simulation results represent time averages of the EEDF-s over 0.5 ns time intervals for $\eta=0$ and over 0.384 ns time intervals for $\eta=10^{-7}$. While the BE solutions are noiseless at any time instance, for consistency, Figs.~\ref{fig:f8}(a) and (b) show BE data that are averaged over the same $\Delta T$ time intervals as in the case of the particle simulations. (Note that the averaged EEDF-s may differ from the instantaneous EEDF-s whenever the distribution function changes non-linearly with time.) The times indicated in Fig.~\ref{fig:f8} correspond to the center of the averaging intervals, except for the initial ($t=0^-$) EEDF-s that correspond to the steady state at $E/N$ = 2.2 Td.

In the absence of electron-electron interactions (Fig.~\ref{fig:f8}(a), $\eta=0$) the EEDF-s obtained from the BE solution and the particle simulations show an excellent agreement. During the relaxation process a peak develops in the EEDF (``inverse-shaped'' EEDF, cf. section 2.2). The position of this peak moves towards lower energies as time proceeds, and at long times the peak disappears. 

In the presence of e-e interactions (Fig.~\ref{fig:f8}(b)), even at the low electron to gas density ratio ($\eta=10^{-7}$) considered, the relaxation of the EEDF proceeds in a significantly different way due to the Maxwellization of the EEDF that results in the (almost complete) disappearance of the peak of the EEDF. This effect acts against the conditions that are needed for the existence of the TNM, as explained in section 2.2. The distribution functions obtained by the particle simulation method and the Boltzmann equation show a good general agreement, although definite differences, which can be attributed to the approximations involved in the ``traditional'' solution of the BE, can be seen in the shapes of the EEDF-s at some moments (especially at early times, including the ``initial'' EEDF that corresponds to $E/N$ = 2.2 Td). Our analysis shows that the value of the drift velocity is not very sensitive to the tail of the EEDF, and that is why the computed drift velocity values shown in Fig.~\ref{fig:f7} agree very well despite of the differences between the EEDF-s seen in Fig.~\ref{fig:f8}.

The final states of the relaxation for both cases investigated were reached only in the BE computations, which show that the steady state EEDF-s at $E/N$~=~0.01~Td are closely Maxwellian with temperatures $T_{\rm e}~\approx$~308~K and 307~K, respectively, for $\eta=0$ and $\eta = 10^{-7}$. The corresponding electron drift velocities are 36.5 m/s and 36.3 m/s, respectively.

The different timescales of the relaxation in the absence and in the presence of e-e interactions can be explained as follows. Recall that the momentum transfer cross section for the electron scattering from Xe atoms has a deep (Ramsauer--Townsend) minimum at energies $\sim$0.6 eV and increases sharply with energy in the energy interval  0.6 eV -- 6 eV. At the time of change of the  electric field ($t$ = 0) there are many electrons in the gas with relatively high energies ($>$2 eV). Electrons of this group lose their energy in elastic collisions, the reduction of the electron energy leads to a fall off of the probability of the elastic scattering and, consequently, to a decrease of the rate of energy losses. If electron-electron collisions are not taken into account, the flux of electrons in energy space from high to low energies results in the formation of the ``inverse shape'' of the distribution function (cf. section 2.2). At $t$ = 49.75 ns (see Fig~\ref{fig:f8}(a)) the majority of the electrons are located (in the energy space) in the interval where the momentum transfer cross section is minimal. For this reason, the further relaxation of the distribution function is rather slow, as shown by the corresponding curves in Fig.~\ref{fig:f7}. In this case, according to the BE calculations, the EEDF reaches the steady state shape at $t \sim$1000~ns. When taken into account, electron-electron collisions prevent formation of the ``inverse shaped'' EEDF and redistribute electrons over a wider energy interval, wherein the momentum transfer cross section is essentially higher than at the R-T minimum ($\approx$ 0.6 eV). As a result, the rate of the electron energy losses increases noticeably and the time of relaxation becomes as short as $\sim$120 ns.  

\section{Conclusions}

In this paper we have investigated two particular effects that appear in electron transport in gases: the Negative Differential Conductivity and the Transient Negative Mobility effects in xenon gas. Computations have been carried out using a first-principles, approximation-free particle simulation technique and using a solution of the Boltzmann transport equation (BE) that included the traditionally adopted approximations, i.e., (i) searched for the distribution function in a two-term form, (ii) neglected the Coulomb part of the collision integral for the anisotropic part of the distribution function, (iii) treated Coulomb collisions as binary events, and (iv) limited the range of the electron-electron interaction to be effective only within a cutoff distance. 

The results obtained from the two methods have been found to be in good qualitative agreement confirming that the BE solutions predict correctly the {\it existence and the qualitative characteristics} of both effects considered here, despite of the approximations adopted in the BE solution. In the case of the NDC effect the differences between the results obtained for $v_{\rm d}$ by the two methods remained below 10\% at 1 atm pressure and have grown up to $\approx$20 \% at 10 atm. For the mean energy, the largest deviations, up to 15\% were also found at the higher pressure. In the case of the TNM effect, the two methods yielded data for the time dependence of the electron drift velocity that agreed within the statistical noise of the simulation. Differences ranging up to a factor of two were found, however, in the high-energy tails of the distribution functions calculated with taking into account e-e collisions. Clarification of the effects of the individual approximations may be aided in the future by the advances in the solutions of the Boltzmann equation, e.g. \cite{Hagelaar}.
\vspace{1cm}

Acknowledgment: the authors thank Prof. A. P. Napartovich for useful discussions and acknowledge the support via the grant OTKA-K-105476.

%

\end{document}